\documentstyle[aps,prl]{revtex}
\draft

\begin{document}
\twocolumn

\title{
A Dynamic Mechanism of Energy Conversion to a Mechanical Work
}

\author{Naoko NAKAGAWA}
\address{
Department of Mathematical Sciences, 
Ibaraki University,
Mito, Ibaraki 310-8512, Japan}

\author{Kunihiko KANEKO}
\address{Department of Pure and Applied Sciences,
College of Arts and Sciences,
University of Tokyo,
Tokyo 153-8902, Japan}

\maketitle

\begin{abstract}

We propose a dynamic mechanism of a molecular machine for energy conversion,
by considering a simple model describing the dynamics of two components,
the head and the chain.
After injection of energy to the head region, the energy
is stored at one part for some time,
and is used step by step, allowing the head to move directionally
along the chain, irrespectively of the direction of the input,
under a fluctuating environment.
Our system can adjust the timing with which the head crosses the energy
barrier by taking advantage of
internal dynamics and the flexibility of components.
The mechanism is shown to be robust and efficient.
Some suggestions are given for molecular machines.

\end{abstract}


Recently, how energy is converted from one form to another form has been
extensively studied.  In biochemical reaction, some molecules are known
to work for ``energy converter". On the other hand, recent progress in
nanotechnology enables us to design a microscopic machine that works as
energy converter. Since
such molecular machine has to function under thermal fluctuations comparable
to the input energy, it is questionable if it works in the same way as
a macroscopic machine.  This question was first addressed up by Oosawa
as a dichotomy
between tight and loose couplings\cite{Oosawa_Hayashi,Oosawa}.  
In the tight coupling, energy is
converted one-to-one, as in our macroscopic machine. In contrast, in the
loose coupling, conversion from chemical energy to mechanical work does
not occur all in once, but step by step.  Here, the output work is not
precisely specified, but is distributed with some fluctuations.

It is important to propose possible mechanisms for the energy conversion
with loose couplings,
since they are relevant to consider a molecular machine working under
thermal fluctuations
that cannot be negligible at that scale.
One of the well-known and successful models for such loose coupling mechanism
is thermal ratchet\cite{Feynmann,Vale_Oosawa,Magnasco,Prost}.  
The model gave one plausible mechanism for the energy conversion
to a mechanical work, and
statistical analysis was carried out successfully.
Here we propose  another possible mechanism, that fully
takes advantage of nonlinear  dynamics of the system in concern.
By taking a Hamiltonian dynamical model,  we can trace each 
process of energy conversion, without using probability distribution,
and study the mechanism in terms of dynamical systems.
Since our model is self contained, we do not have to assume external
switch of potential or specific non-white noise.

In general, we are interested in a system composing of several degrees of freedom.  
The system has an `input part' to which energy is injected, and an `output part'
from which a mechanical work is extracted. Temporal evolution of the
system is given by a set of equations with Hamiltonian dynamics with
damping and noise coming from a heat bath. 
Since the system has several
degrees of freedom, the motion generally generates fluctuations from
their chaotic dynamics in addition to those by heat bath.  
Relevance of Hamiltonian dynamics to a rectifier is
also presented by Terraneo et al\cite{Peyrard_Casati}.

Here we demonstrate that mechanical work is extracted as a directional motion, 
among all the degrees of the system, even though
input is not specifically controlled with regards to its timing
or direction, and without any `supervisor' (i.e., external control).
The model is capable of functioning with the injection of
energy that is slightly larger than the energy associated
with the thermal fluctuations in the system.

The model system consists of a motor that interacts with a chain
composed of $N$-lattice sites, positioned at $x_i$ with index $i$.
The motor consists of a `head' of position $x_h$
and one internal degree of freedom
in the form of a `pendulum' represented by $\theta$.
The injection of energy into the system
is represented by the transfer of energy to this pendulum. 
The interaction potential $V(x_h-x_i,\theta)$
between the chain and the head is spatially asymmetric and its form
depends on the angle of the pendulum (see Fig.1).
The periodic lattice is assumed in the chain,
to consider directional motion in an asymmetric periodic potential 
as is often studied in the study of thermal ratchet.
Every degree of freedom, except for the internal pendulum, 
is in contact with a heat bath,
generating random fluctuations described by a Langevin equation
with damping.
The equations of motion for this system are chosen as
\begin{eqnarray}
m_c \ddot x_i &=& -m_c \gamma \dot x_i +\sqrt{2m_c \gamma T}\xi_i(t)\nonumber\\
&-&K_c\left\{ (x_i-iL)+(x_i-\frac{x_{i-1}+x_{i+1}}{2})\right\}
-\frac{\partial V}{\partial x_i}, \nonumber\\
m_h \ddot x_h &=& -m_h \gamma \dot x_h +\sqrt{2m_h \gamma T}\xi_h(t)
-\sum_i\frac{\partial V}{\partial x_h}, \nonumber\\
m_{\theta} \ddot \theta &=& -\sum_i\frac{\partial V}{\partial \theta},\nonumber
\end{eqnarray}
where $T$ is the temperature, $\gamma$ is a friction coefficient 
and $\xi_{\alpha}(t)$ represents Gaussian white noise.
Here, we use the units Boltzmann constant $k_B=1$.
$K_c$ and $L$ are the spring constant and the natural interval 
between two neighboring lattice sites in the chain.
$m_c$, $m_h$ and $m_\theta$ are mass of the respective degrees of freedom.

The potential form is asymmetric in space as shown in Fig.1,
where the characteristic decay length of the interaction is set 
at a smaller value than $L$, 
to assure that the interaction is
confined mostly to the nearest lattice sites. 
In this Letter, we adopt the following potential form,
$$
V(\Delta x,\theta)= K_h\frac{\tanh(p \Delta x -r)
+(1-\cos\theta)/2}{ \cosh(\Delta x/d)},
$$
where the parameters $p$ and $r$ determine the degree of asymmetry
and $K_h$ and $d$ gives a strength and decay length of interaction, 
respectively.
Specific choice of this form is not important.
We have simulated our model choosing several other potential forms with asymmetry
and obtained qualitatively the same results for the directional motion.

In thermal equilibrium, no directional motion is possible on the average.
However, when energy is imported to the pendulum,
directional motion is observed, after a time lag.
One realization of time series of the positions of the head,
pendulum, and neighboring lattice
sites is plotted in Fig. 2.  Although this time series represents
a typical example of the head motion,
the particular motion differs for each event, since it includes
a chaotic component and is subject to thermal noise.
We computed the distribution of the number of steps 
as displayed in Fig.3.  In this distribution, there appears
a peak at two steps, which
is shifted to larger values as the amount of injected energy is increased.
The distribution was obtained from $1000$ random sequences
$\xi_{\alpha}(t)$, where, at the energy injection, configuration of the system
and the direction of the rotation of the internal pendulum
are not taken into account.
The direction of the head motion is independent of the direction of the rotation
of the pendulum and other
situation of the system.  Therefore, the system functions robustly.

The presently studied energy conversion mechanism functions over a rather broad
range of temperatures, as shown in Fig.4(a).
The average number of steps is a linearly increasing function
of the injected energy above a certain threshold value.
This implies that as the amount of injected energy increases,
the efficiency of the energy conversion does not decay to zero,
but, rather, approaches some finite constant.
For low temperatures, there exists a critical value of $E_0$
around $\delta V=0.34$ (see Fig.1),
below which the directional motion is suppressed.
For higher temperatures, the head exhibits directional motion
even for small values of $E_0$.
It is thus seen that the system's directional motion is robust
with respect to changes of the environmental temperature and can adjust itself
in response to the amount of input energy.
To make this energy conversion possible, it is important
that the chain be sufficiently flexible.  Indeed, if the
spring constant $K_c$ is too large, no directional motion
of the head is generated (see Fig.4(b)).

Now, we discuss how energy conversion is carried out, by closely examining the
dynamics of the model.  As depicted in Fig.2, when the pendulum motion is
fast, and is essentially decoupled from the head motion,
the head and the nearest lattice site exhibit highly correlated
vibration (stage {\bf a}).  
Since their motion is governed mainly by a two-body interaction,
this coherent vibration is not surprising.
Here the energy is stored at the pendulum for some time, as was already studied
\cite{Nakagawa_Kaneko}, in possible
relationship with the Boltzmann-Jeans conjecture\cite{BJ}.

As an initially excited pendulum relaxes, their motion is eventually
no longer decoupled from the rest of the system, and the pendulum, head, and
the corresponding lattice site come to exhibit three-body motion.
Since this motion possesses instability, the coherent
motion of the head and the lattice site is lost.
As a result, the vibration amplitude of the head becomes larger
with energy transfer from the pendulum (stage {\bf b}).
The energy is localized at
the corresponding lattice site and the head for a certain interval
(see $T_0$ and $T_h$).
Then, due to the large amplitude, the head begins to interact with the neighboring lattice sites.
As it does so, head begins to experience the asymmetry of the potential,
and its motion becomes synchronized with that of the neighboring site
immediately to the left of the original (stage {\bf c}).
Because the interaction between the head and the left lattice site is repulsive
(see $V(x_h-x_i)$ in $x_h-x_i>L/2$), the head
bounces back eventually to the neighboring site immediately to the right of the original
after a few period of synchronization (stage {\bf d}).  In this process, the
head absorbs energy from the chain, so that it can cross over the
energy barrier to the next site to the right.
The timing of this crossing is spontaneously determined by
the interaction of the head, pendulum, and lattice sites.

If the injected energy were dissipated completely into
a large number of degrees of freedom (as `heat'),
the conversion of energy into mechanical work
would suffer a rather large loss.
Contrastingly, the motion generated by our mechanism remains
confined to just a few degrees of freedom and
highly correlated in space and time.
In other words, supplied energy to the pendulum is not diffused
randomly as heat, and, as a result, the conversion is efficient.
For the robustness of this conversion mechanism, 
it is important that the motion is not periodic but chaotic.
Since chaotic motion allows continuous spectrum of frequency,
in contrast to resonant periodic motion,
flexible adaptation of the energy conversion mechanism to changes of the
governing parameters is possible. 
Hence, neither fine-tuning of parameters nor external control for a driving force
is needed.

This is in strong contrast with thermal ratchet models, 
popular in the theoretical study of molecular motors
\cite{Feynmann,Vale_Oosawa,Magnasco,Prost}.
In such models, to attain reasonable efficiency, it is necessary to
assume a rather fine-tuned timing for the switching from one
potential form of the interaction to another,
or to tune the time scale of the colored noise present.
When there is such fine-tuning, the motion of the ratchet and this
switching are synchronized, and the energy conversion  is no longer loose.
Furthermore, either the switching or the specific form of the noise
is externally given.
In the present model, such external condition required for the thermal ratchet
is generated through the dynamics.
In addition, in the present model we can trace each event of
the conversion of the energy to mechanical work,
as in  recent single-molecule experiments.

An advantage of using Hamiltonian dynamics is that in this case
the energy transfer can be traced directly, as shown in Fig.2.
We have found that for the present mechanism to function
energy must be localized within the head and neighboring lattice sites.  
This condition is not satisfied, for example, when
the spring is stiff (i.e., when $K_c$ is large).
In that case, energy rapidly diffuses from the head to all lattice sites,
and the conversion mechanism does not work, as shown in Fig.4(b).
The localization
of energy in the few lattice sites near the head over some time interval allows
for efficient conversion, which results in directional motion.

We have demonstrated how a coupled dynamical system
with an internal degree of freedom
can carry out robust conversion of energy to mechanical work
generating directional motion.
Note that the efficiency of our mechanism is lost
when the spring of the chain is stiff as shown in Fig.4(b).
This is in strong contrast with a consequence from the tight machine,
where the efficiency should be higher as the chain is stiffer.

Although we have demonstrated the behavior of our mechanism
only in a weakly dissipative system, this type of
mechanism is common in dynamical systems, and proper extension of
the present study to over-damped systems should be possible.
Such an extension would be useful in constructing a
more realistic model of a biomotor based on the present mechanism.
It is interesting to note that the broad distribution form of steps
in the present model is similar to that
observed in some experiments of molecular motors\cite{Kitamura_Tokunaga},
where very large time-lag between the injection of the energy and the
directional motion is also observed\cite{Ishijima_Yanagida}.

Because the mechanism studied here operates in a Hamiltonian system,
it represents a possible type of energy conversion mechanism
at a molecular scale. 
It would be interesting
to study enzyme function using this mechanism.
Furthermore, it seems that designing
a nano-machine for energy conversion employing our mechanism
would not be too difficult, because this mechanism functions in the presence of
thermal fluctuations corresponding to energies that are of the same order
as the injected energy.


Acknowledgment:
We thank K. Kitamura and T. Yomo for critical reading of this manuscript.
The present work is supported by
Grants-in-Aids for Scientific Rsearch from
the Ministry of Education, Science and Culture of Japan.

\begin{figure}
\caption{
Profile of the model. 
The form of $V$ depends on the value of $\theta$,
where, the solid curve represents the form for $\theta=0$
(the equilibrium state for $\theta$),
and the dotted curve for $\theta=\pi$.
The latter value appears upon excitation,
consisting of the instantaneous increase to $E_0$ of the kinetic energy
for the pendulum. 
In this Letter, the following parameters are fixed 
as $m_c=m_h=1$, $m_\theta=0.02$ and $L=1$.
The chain consists of $40$ lattice sites with periodic boundary condition.
$K_c$ is used as a control parameter.
$\gamma$ is $0.01$. 
The parameters for the potential $V(\Delta x, \theta)$ (see the text) are 
$p=10$, $r=3$, $K_h=0.2$ and $d=L/4$. 
}
\end{figure}

\begin{figure}
\caption{
A typical relaxation process following excitation of the pendulum.
$K_c=0.5$.
In the simulation, the system was prepared in thermal equilibrium
with $T=0.02$,
and the pendulum was excited at $time=0$ with $E_0=0.4$.
With the temperature used here, the head remains
at one lattice site for a very long time in thermal equilibrium.
Upper: Time series of the positions of the head $x_h$ (red)
and a few neighboring lattice sites $x_i$ ($-1\le i\le 3$, green).
Middle: Time series of the kinetic energy of the internal pendulum
($T_{\theta}$),
the lattice sites ($T_i$) and the head ($T_h$), where the vertical scale
of $T_{\theta}$ is larger than that of the others.
In the lower schematic figures,
{\bf a - d} correspond to each stage shown in the upper figure.
}
\end{figure}

\begin{figure}
\caption{
Frequency distribution of the displacement (number of steps)
of the head position $x_h$ per excitation.
$K_c=0.5$, $T=0.02$ and $E_0=0.4$ ($=20k_BT$).
}
\end{figure}

\begin{figure}
\caption{
(a) Average displacement $\langle \Delta x_h\rangle$ as a function of 
$E_0$ for three values of temperature,
computed as the ensemble average over 1000 samples. $K_c=0.5$.
For larger values of $E_0$,
$\langle \Delta x_h\rangle$ increases as a function of $E_0$ 
in similar manners for all $T$. 
(b) Dependence of $\langle \Delta x_h\rangle$ on $K_c$
for $T=0.02$ and $E_0=0.35$.
Directional motion is most prominent in a particular range of
stiffness of the chain ($K_c \approx 0.5$), where
the fluctuations of the lattice are slightly larger than those of the head.
The directional motion is suppressed both for larger values of $K_c$,
because lattice fluctuations
decrease in magnitude, and for smaller values of $K_c$
because the head interacts not only with the lattice site 
at which it is positioned, but also with
neighboring sites, resulting in an effectively
stronger potential experienced by the head.
}

\end{figure}

\end{document}